\documentclass[conference]{IEEEtran}
\usepackage{graphicx}
\usepackage{epsf}
\usepackage{amsmath, amsfonts}
\usepackage{latexsym}
\usepackage{subfigure}
\usepackage{multirow}
\usepackage{multicol}
\usepackage[table]{xcolor}
\usepackage{capt-of}

\begin{document}

\title{\fontsize{24}{10}\selectfont Human Exposure to RF Fields in 5G Downlink}

\author
{
Imtiaz Nasim and Seungmo Kim\\
\fontsize{10}{10}\selectfont \{in00206, seungmokim\}@georgiasouthern.edu\\
\fontsize{10}{10}\selectfont Department of Electrical Engineering, Georgia Southern University\\
\fontsize{10}{10}\selectfont Statesboro, GA 30460, USA
}

\maketitle

\begin{abstract}
While cellular communications in millimeter wave (mmW) bands have been attracting significant research interest, their potential harmful impacts on human health are not as significantly studied. Prior research on human exposure to radio frequency (RF) fields in a cellular communications system has been focused on uplink only due to the closer physical contact of a transmitter to a human body. However, this paper claims the necessity of thorough investigation on human exposure to downlink RF fields, as cellular systems deployed in mmW bands will entail (i) deployment of more transmitters due to smaller cell size and (ii) higher concentration of RF energy using a highly directional antenna. In this paper, we present human RF exposure levels in downlink of a Fifth Generation Wireless Systems (5G). Our results show that 5G downlink RF fields generate significantly higher power density (PD) and specific absorption rate (SAR) than a current cellular system. This paper also shows that SAR should also be taken into account for determining human RF exposure in the mmW downlink.
\end{abstract}

\vspace{0.2 in}

\begin{IEEEkeywords}
5G; mmW; Downlink; Human RF exposure; PD; SAR.
\end{IEEEkeywords}

\IEEEpeerreviewmaketitle

\begin{table*}[t]
\centering
\begin{minipage}{0.55\textwidth}
\scriptsize
\caption{Parameters for 5G and Release 9}
\begin{tabular}{|c|c|c|}
\hline
\textbf{Parameter} & \multicolumn{2}{|c|}{\textbf{Value}}\\ \hline \hline
&\multicolumn{1}{|c|}{\cellcolor{gray!10}5G} & \multicolumn{1}{|c|}{\cellcolor{gray!10}Release 9}\\ \hline
Carrier frequency & {28 GHz} & 1.9 GHz\\
System layout & {RMa, UMa, UMi \cite{tr38900}} & {SMa, UMa, UMi \cite{ts25996}}\\
Inter-site distance (ISD) & {200 m} & {1,000 m}\\
Cell sectorization & {3 sectors/site} & 6 sectors/site\\
Bandwidth & {850 MHz} & 20 MHz \\
Max antenna gain & {5 dBi per element} & 17 dBi\\
Transmit power & 21 dBm per element & 43 dBm\\
AP's number of antennas ($\lambda/2$ array) & 8$\times$8 and 16$\times$16 & 4$\times$4\\
AP antenna height & 10 m & 32 m\\ \hline
Duplexing & \multicolumn{2}{|c|}{Time-division duplexing (TDD)} \\
Transmission scheme & \multicolumn{2}{|c|}{Singler-user (SU)-MIMO} \\
UE noise figure & \multicolumn{2}{|c|}{7 dB}\\
Temperature & \multicolumn{2}{|c|}{290 K}\\ \hline
\end{tabular}
\label{table_parameters}
\end{minipage}\hspace{0.1in}
\begin{minipage}{0.4\textwidth}
\centering
\includegraphics[width = \linewidth]{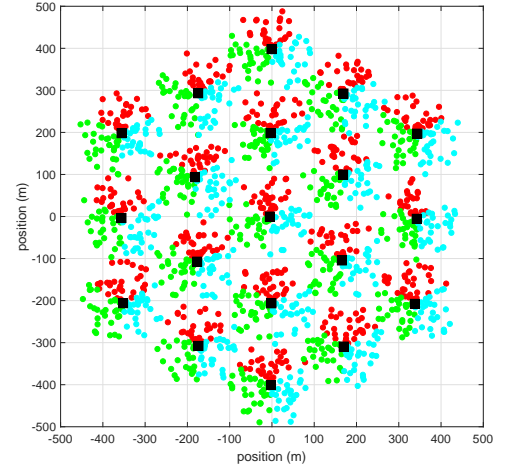}
\captionof{figure}{A snapshot of one ``drop'' of 5G topology (19 sites, 3 sectors per site, and 30 UEs per sector)}
\label{fig_topology}
\end{minipage}
\end{table*}

\section{Introduction}\label{sec_intro}
It is acknowledged that exposure to RF has negative impacts on human body. The rapid proliferation of mobile telecommunications has occurred amidst controversy over whether the technology poses a risk to human health \cite{wu15}. At mmW frequencies where future mobile telecommunications systems will likely operate, two changes that will likely occur have the potential to increase the concern on exposure of human users to RF fields. First, \textit{larger numbers of transmitters} will operate. More base stations (BSs) will be deployed due to proliferation of small cells \cite{saadeh17}-\cite{agiwal16} and mobile devices accordingly. This will increase chance of human exposure to RF fields. Second, \textit{narrower beams} will be used as a solution for the higher attenuation in higher frequency bands \cite{rappaport13}-\cite{akendiz14}. Very small wavelengths of mmW signals combined with advances in RF circuits enable very large numbers of miniaturized antennas. These multiple antenna systems can be used to form very high gains. Such higher concentration of RF energy will increase the potential to more deeply penetrate into a human body.

\subsection{Related Work}\label{sec_related}
This paper is motivated from the fact that prior work is not enough to address such potential increase in threats.

\subsubsection{Measurement of Human RF Exposure}
Being aware of the health hazards due to electromagnetic (EM) emissions in mmW spectrum, international agencies such as the Federal Communications Commission (FCC) \cite{fcc01} or the International Commission on Non-Ionizing Radiation Protection (ICNIRP) \cite{icnirp98} set the maximum radiation allowed to be introduced in the human body without causing any health concern. Possibilities of skin cancer due to RF emissions at higher frequency spectrum are reported \cite{gao12}. Heating due to EM exposure in mmW is absorbed within the first few millimeters (mm) within the human skin; for instance, the heat is absorbed within 0.41 mm for 42.5 GHz \cite{alekseev01}. The mmW induced burns are more likely to be conventional burns as like as a person touching a hot object as reported in \cite{wu15}. The normal temperature for the skin outer surface is typically around 30 to 35$^{\circ}$C. The pain detection threshold temperature for human skin is approximately 43$^{\circ}$C as reported and any temperature over that limit can produce long-term injuries. 

One problem is that the literature on the impact of cellular communications on human health is not mature enough. The three major quantities used to measure the intensity and effects of RF exposure are SAR, PD, and the steady state or transient temperature \cite{em92}\cite{em05}. However, selection of an appropriate metric evaluating the human RF exposure still remains controversial. The FCC suggests PD as a metric measuring the human exposure to RF fields generated by devices operating at frequencies higher than 6 GHz \cite{fcc01}, whereas a recent study suggested that the PD standard is not efficient to determine the health issues especially when devices are operating very close to human body in mmW \cite{rappaport15}. Therefore, this paper examines the human RF exposure by using both PD and SAR.

\subsubsection{Reduction of Human RF Exposure}
Very few prior studies in the literature paid attention to human RF exposure in communications systems \cite{wu15}\cite{rappaport15}-\cite{chahat12}. Propagation characteristics at different mmW bands and their thermal effects were investigated for discussion on health effects of RF exposure in mmW radiation \cite{rappaport15}. Emission reduction scheme and models for SAR exposure constraints are studied in recent work \cite{sambo15}\cite{love16}.

However, health impacts of mmW RF emissions in \textit{downlink} of a cellular communications system have not been studied so far, which this paper targets to discuss.

\subsection{Contributions}
Three contributions of this paper can be highlighted and distinguished from the prior art.

Firstly, this paper analyzes the human RF exposure in the \textit{downlink}. All the prior work studied an uplink only, while paid almost no attention to suppression of RF fields generated by access points (APs) and BSs in a 5G nor Release 9 network, respectively. In fact, APs generate even stronger RF fields compared to the concurrent systems, due to (i) higher transmit power and (ii) larger antenna array size leading to higher concentration of RF energy. Moreover, one important feature of the future cellular networks is small cell networks. The consequences of this change will be two-fold: (i) APs/BSs will serve smaller geographic areas and thus are located closer to human users; (ii) larger numbers of APs/BSs will be deployed, which will lead to higher chances of human exposure to the RF fields generated by downlinks.

Secondly, this paper finds that \textit{SAR should also be considered} in determination of human RF exposure in mmW downlinks. Our simulations are performed for a 5G system based on the 3GPP Release 14 \cite{tr38900}, one of the promising technical specifications for 5G. The results show that even considering a shallow penetration into a human body due to high frequencies, a downlink RF emission causes significantly higher SAR in mmW. This effectively highlights the elevation in potential harmful impact in human health, which can ignite higher interest in further research on design of future cellular communications systems considering the impacts on human RF exposure.

Thirdly, it explicitly \textit{compares the human RF exposure in downlinks between 5G and Release 9}, highlighting the difference in the size of a cell. This will lead to clear understanding on how the technical evolution to 5G affects the human RF exposure. This paper calculates PD and SAR of a 5G \cite{tr38900} and a Release 9 \cite{ts25996} to highlight the change in human RF exposure according to the technical evolution.


\section{System Model}
This section describes the system setting for a cellular communications network that forms the basis for the analysis of human RF exposure. Considering the frequency spectrum of 28 GHz as a potential candidate for 5G, we use a corresponding technical report \cite{tr38900} that was released by the 3GPP. Also, this paper compares the human RF exposure level in a 5G system to a legacy cellular communications system. For highlighting how much a SAR level can be increased compared to the current wireless services, this paper chose to compare the 5G to the Release 9 \cite{ts25996}. The parameters of both systems are summarized in Table \ref{table_parameters}.

\subsection{5G}
\subsubsection{Path Loss}
Our model for a 5G system is illustrated in Fig. \ref{fig_topology}. It consists of 19 sites each having 3 sectors. The inter-site distance (ISD) is 200 meters (m) and each sector is assumed to have 30 active user equipments (UEs). Also, as identified in Table \ref{table_parameters}, for the terrestrial propagation between an AP and a UE, the following three path loss models are assumed: Rural Macro (RMa), Urban Macro (UMa), and Urban Micro (UMi) \cite{tr38900}.

\subsubsection{Antenna Beam Pattern}
For a 5G AP, the attenuation patterns of an antenna element on the elevation and azimuth plane are given by \cite{tr38900}
\begin{align}
\label{eq_antenna_5g_a}
A_{a}\left(\phi\right) &= \min\left\{12\left(\frac{\phi}{\phi_{3db}}\right)^2, A_m\right\} \rm{~[dB]}\\
\label{eq_antenna_5g_e}
A_{e}\left(\theta\right) &= \min\left\{12\left(\frac{\theta-90^{\circ}}{\theta_{3db}}\right)^2, A_m\right\} \rm{~[dB]}
\end{align}
where $\phi$ and $\theta$ are angles of a beam on the azimuth and elevation plane, respectively; $\left(\cdot\right)_{3db}$ denotes an angle at which a 3-dB loss occurs. Then the antenna element pattern that is combined in the two planes is given by
\begin{align}\label{eq_antenna}
A\left(\theta,\phi\right) = \min \left(A_{a}\left(\phi\right)+A_{e}\left(\theta\right), A_m\right)\rm{~[dB]}
\end{align}
where $A_m$ is a maximum attenuation (front-to-back ratio). It is defined $A_m = 30$ dB in \cite{tr38900}, but it can be higher in practice. Finally, an antenna gain that is formulated as
\begin{align}\label{eq_geometry_G_final}
G\left(\phi,\theta\right)=G_{max} - A\left(\phi,\theta\right) \rm{~[dB]}
\end{align}
where $G_{max}$ is a maximum antenna gain.

\subsection{Release 9}
\subsubsection{Path Loss}
A cellular network operating on Release 9 is designed to form a cell radius of 500 m, which results in an ISD of 1,000 m. This paper calculates the received power in a downlink, following the path loss models provided in \cite{ts25996}--Suburban Macro (SMa), UMa, and UMi.

\subsubsection{Antenna Beam Pattern}
The antenna radiation pattern for a Release 9 BS is also given as (\ref{eq_antenna_5g_a}) and (\ref{eq_antenna_5g_e}). However, unlike at a 5G AP, $\theta_{3db}$ and $A_m$ for a Release 9 BS are given as 35$^\circ$ and 23 dB, respectively.

\begin{figure*}
\minipage{0.45\textwidth}
\centering
\includegraphics[width = \linewidth]{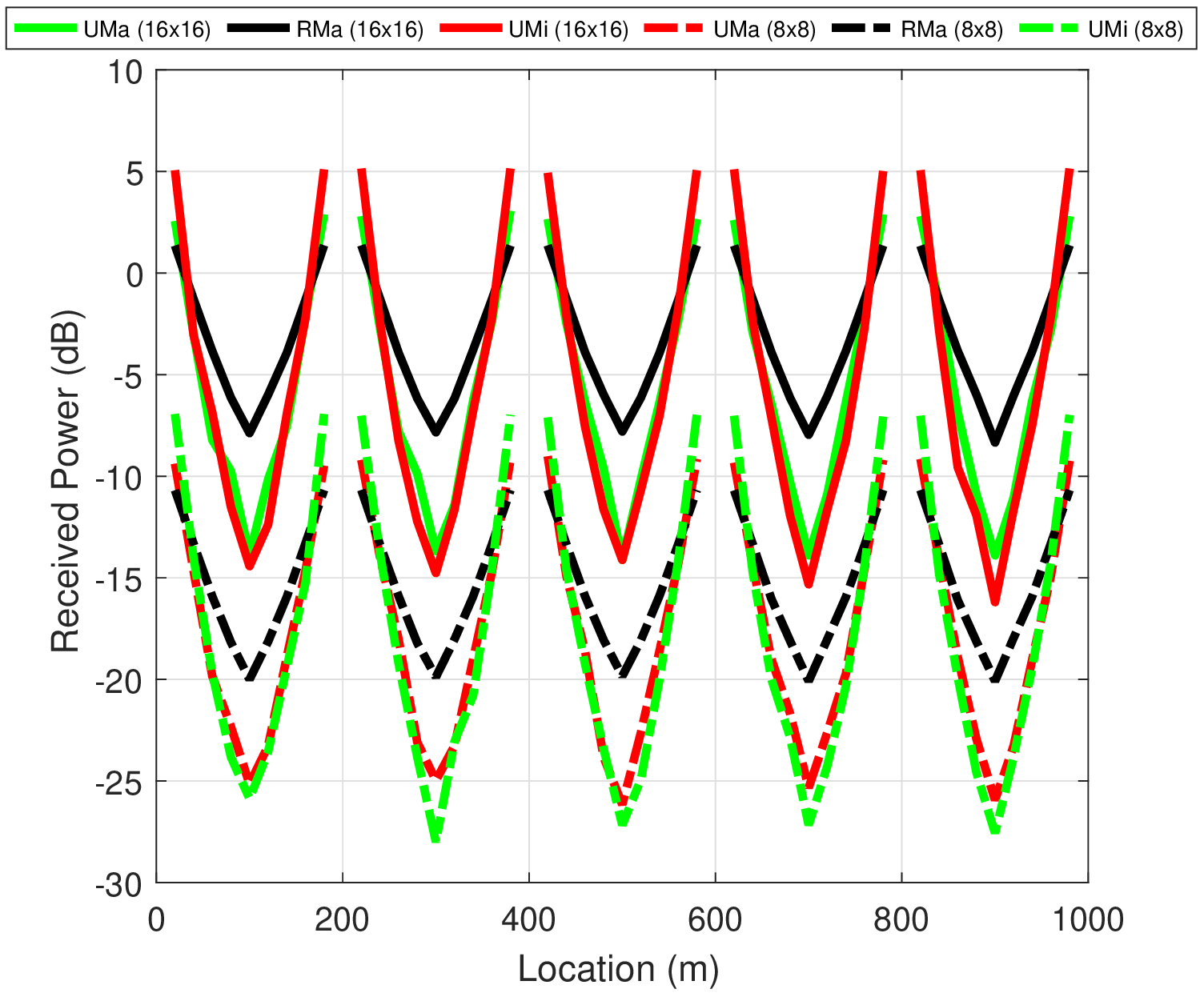}
\caption{Received signal power (\ref{eq_pr}) versus UE location in a 5G system (APs are located at 0, 200, 400, 600, 800, and 1,000 m)}
\label{fig_pr_5g}
\endminipage\hfill
\minipage{0.45\textwidth}
\centering
\includegraphics[width = \linewidth]{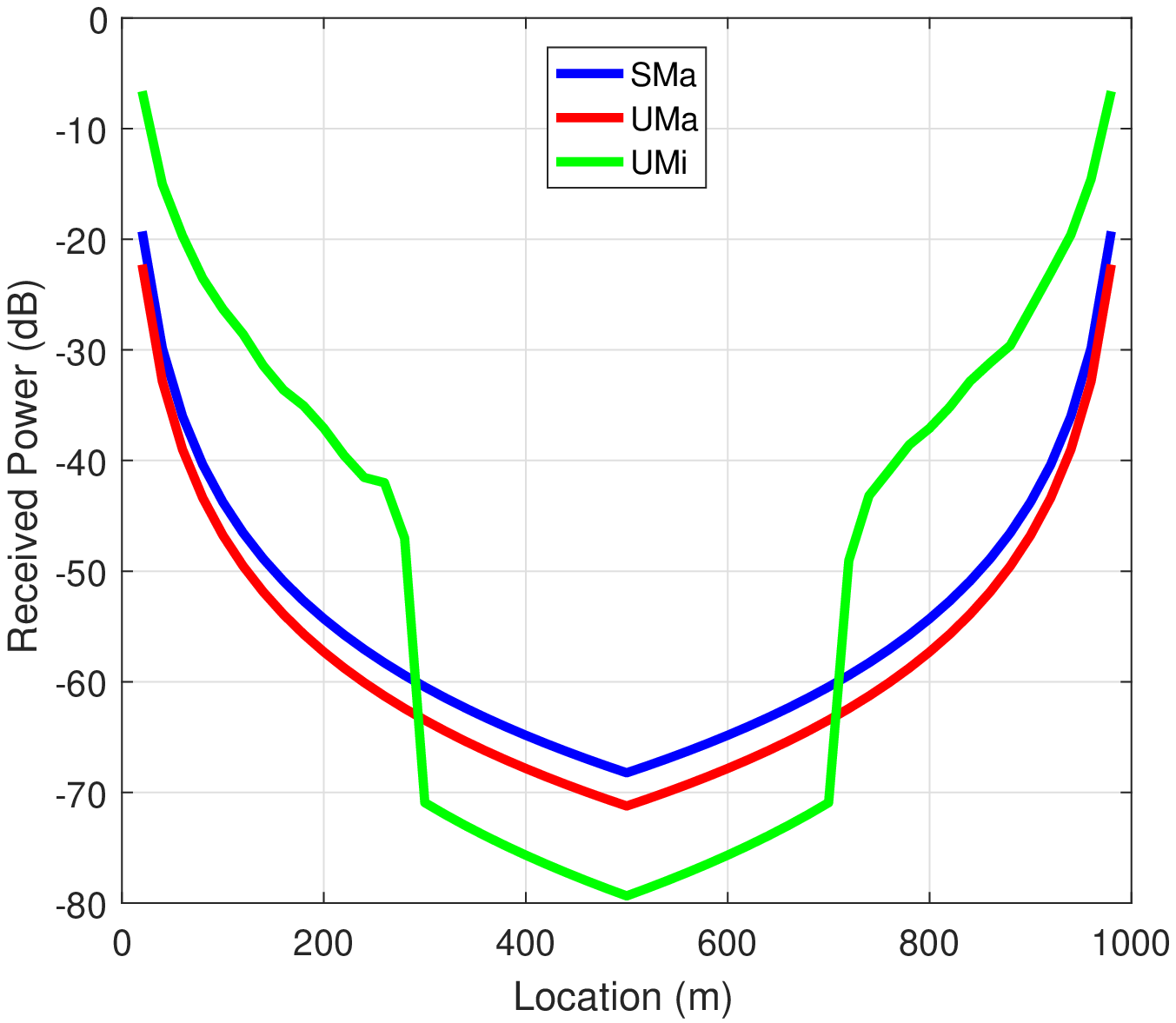}
\caption{Received signal power (\ref{eq_pr}) versus UE location in a Release 9 system (BSs are located at 0 and 1,000 m)}
\label{fig_pr_release9}
\endminipage
\end{figure*}

\section{Performance Analysis}
In this section, we present an analysis on the human RF exposure in a 5G communications and a Release 9 system. Though we chose 28 GHz frequency spectrum for 5G performance analysis, performance for any other frequency spectrum can be demonstrated following the same methodology. It is obvious that the higher number of elements used in the antenna give better signal power, the outcome also increases the cost and complication of the antenna design. The present technology has a large cell size where a single BS can provide coverage to more than thousands of meters, but the cell size of 5G is relatively small. In a model like Release 9, there may be one BS used to provide coverage to a wide area for providing service to UEs, but in 5G scenario, the same area is covered by a number of scattered APs to provide a better reliable service.

\subsection{Data Rate}
The downlink performance of a system is calculated from the Shannon's formula, which is given by
\begin{align}\label{eq_r}
R = B \log(1+\rm{SNR})
\end{align}
where $R$ and $B$ denotes a data rate and bandwidth, respectively. Signal-to-noise power ratio (SNR) is used to determine a data rate. Note that the inter-cell interference is not considered for simplicity in calculation as the focus of this paper is analysis of human exposure level, which is not influenced by the interference. In this paper, we calculate a SNR for the UEs considering all the possible locations in a sector that is formed by an AP in a 5G system and a BS in a Release 9 system. However, an accurate \textit{three-dimensional distance} is considered with the exact heights of an AP, BS, and UE which are taken into account referred from \cite{tr38900}. In other words, although the horizontal axes of the results provided in Section \ref{sec_results} present all the possible locations in a cellular system, they in fact demonstrate three-dimensional distances with the exact vertical distances accounted.

The core part in calculation of a SNR is a received power that is directly determined by a path loss model provided in the specifications \cite{tr38900}\cite{ts25996}. Here we provide an analysis framework for the signal power that is received by a UE from either an AP or a BS in a single downlink, denoted by $P_{R,ue}$. It is noteworthy that with straightforward modifications, this framework can easily be extended to an uplink received signal power also. A received signal strength in a downlink transmission of a single sector is computed by averaging over all possible downlink directions according to position of the UE, which is given by
\begin{align}\label{eq_pr}
&P_{R,ue}\left(\mathtt{x}_{ue}\right)\nonumber\\
&= \frac{1}{\left|\mathcal{R}_k^2\right|} \int_{\mathtt{x}_{ue}^{\left(k\right)} \in \mathcal{R}_k^2} \frac{P_{T,ap}G_{ap}\left(\mathtt{x}_{ue}\right)G_{ue}\left(\mathtt{x}_{ue}\right)}{PL_{ap \rightarrow ue}} d\mathtt{x}_{ue}
\end{align}
where $\mathcal{R}_k^2$ is region of a sector and thus $\left|\mathcal{R}_k^2\right|$ is the area of a sector; $\mathtt{x}_{ue}$ is position of a UE in an $\mathcal{R}_k^2$; $P_{T,ap}$ is transmit power of an AP; $G_{ap}$ and $G_{ue}$ are the antenna beamforming gains of an AP and a UE, respectively, in a downlink transmission based on (\ref{eq_geometry_G_final}); $PL_{ap \rightarrow ss}$ is the path loss between the AP and the UE.

\subsection{Human RF Exposure}
To determine the deleterious impacts of RF emissions to the human body in mmW spectrum, SAR and PD are the most commonly used evaluation criteria so far. As there remains a controversy which method is more accurate one to be considered, whether it is a far-field or near-field case, we show both the analysis for SAR and PD for future technology.

The SAR is a quantitative measure that represents the power dissipated per body mass. It is one of the International System of Units (SI), which is measured in watts (W) per kilogram (kg) and is given by
\begin{align}\label{eq_sar}
\text{SAR} = \frac{P_{diss}}{m} = \frac{\sigma \left|E\right|^2}{\rho}
\end{align}
where $P_{diss}$ represents dissipated power in tissue in the unit of W, $m$ represents the exposed tissue mass in the unit of kg, $\rho$ is the tissue mass density (kg/m$^3$), $\sigma$ is the conductivity in siemens per meter (S/m) and $E$ is a root mean square (rms) value of the electric-field strength which is given in the unit of voltage per meter (V/m). The SAR for a particular tissue in human body is different from the SAR for a tissue at different location. Also, SAR at the surface of the exposed tissue is different from the SAR deep within that exposed tissue.

The PD of a transmitting antenna for the far-field can be expressed as \cite{wu15}
\begin{align}\label{eq_pd}
\text{PD} = \frac{\left|E_i\right|^2}{\eta} = \frac{\eta}{\left|H_i\right|^2}
\end{align}
where $E_i$ (V/m) and $H_i$ (A/m) are rms values of the electric and magnetic field strengths, respectively, incident on the tissue surface and $\eta$ is the wave impedance in the unit of ohm ($\Omega$). The SI unit of a PD is W/m$^2$, which indicates that a PD is a measurement of the power dissipated per area of the exposed tissue.

Our paper focuses on the downlink behaviors when performing the analysis and comparison of the two communications system. Incident PD for far-field communications is expressed as
\begin{align}
S_i = \frac{P_T G_T}{4 \pi d^2}
\end{align}
where $P_T$ is a transmit power; $G_T$ is a transmit antenna gain; $d$ is the AP-UE distance (m) as in (\ref{eq_pr}). 

\begin{figure*}
\minipage{0.45\textwidth}
\centering
\includegraphics[width = \linewidth]{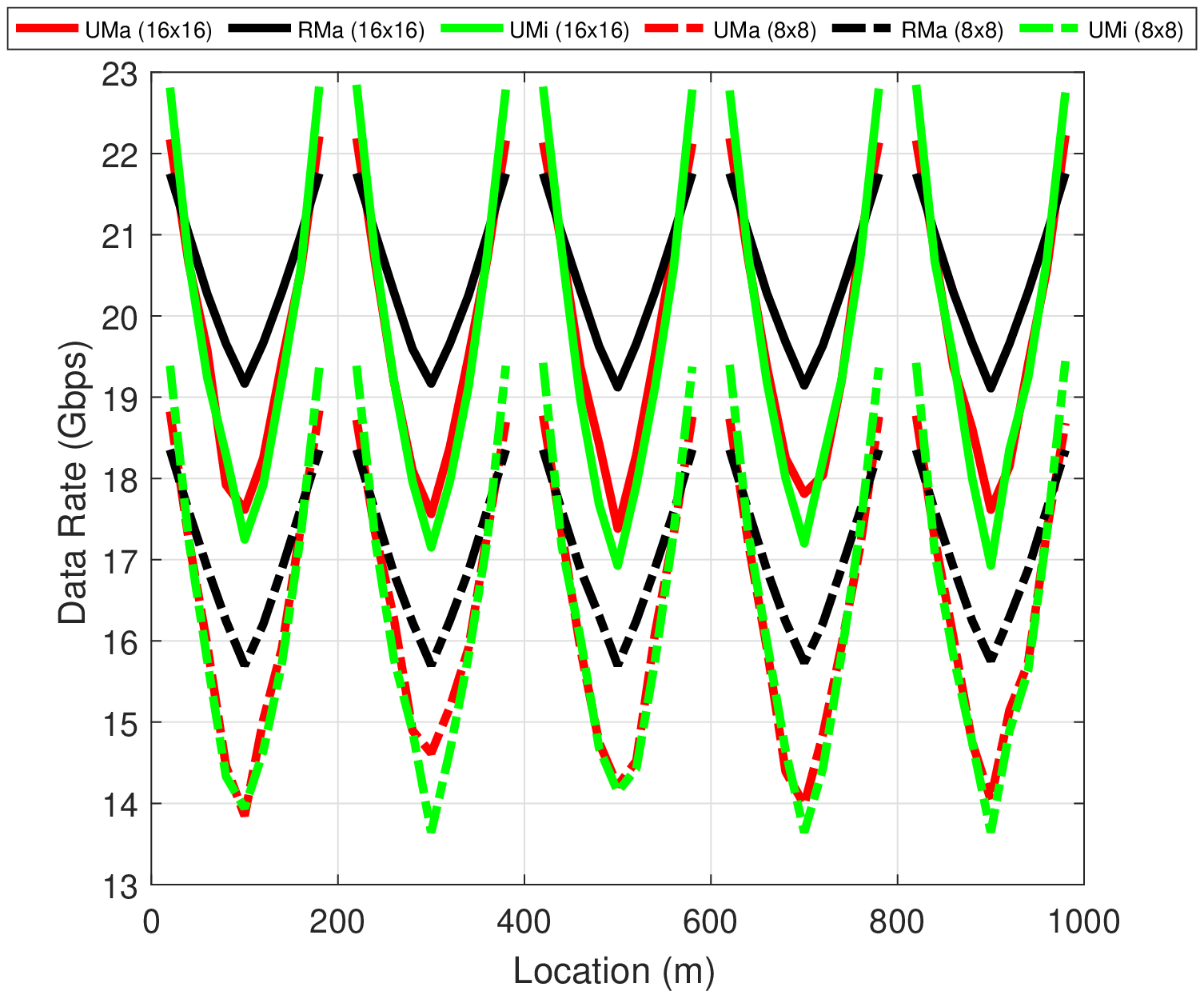}
\caption{Data rate (\ref{eq_r}) versus UE location in a 5G system (APs are located at 0, 200, 400, 600, 800, and 1,000 m)}
\label{fig_rate_5g}
\endminipage\hfill
\minipage{0.45\textwidth}
\centering
\includegraphics[width = \linewidth]{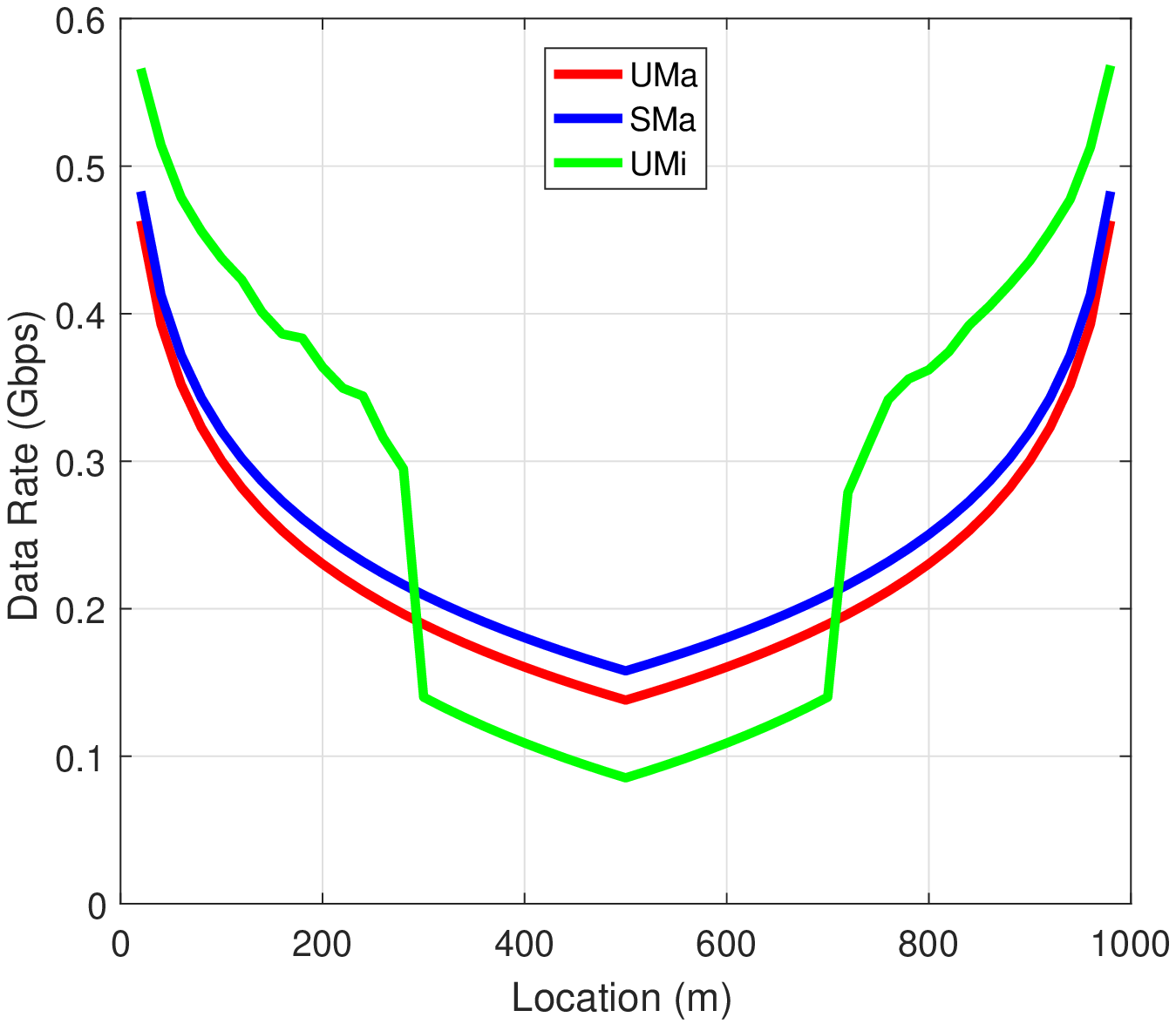}
\caption{Data rate (\ref{eq_r}) versus UE location in a Release 9 system (BSs are located at 0 and 1,000 m)}
\label{fig_rate_release9}
\endminipage
\end{figure*}

\begin{figure*}
\minipage{0.45\textwidth}
\centering
\includegraphics[width = \linewidth]{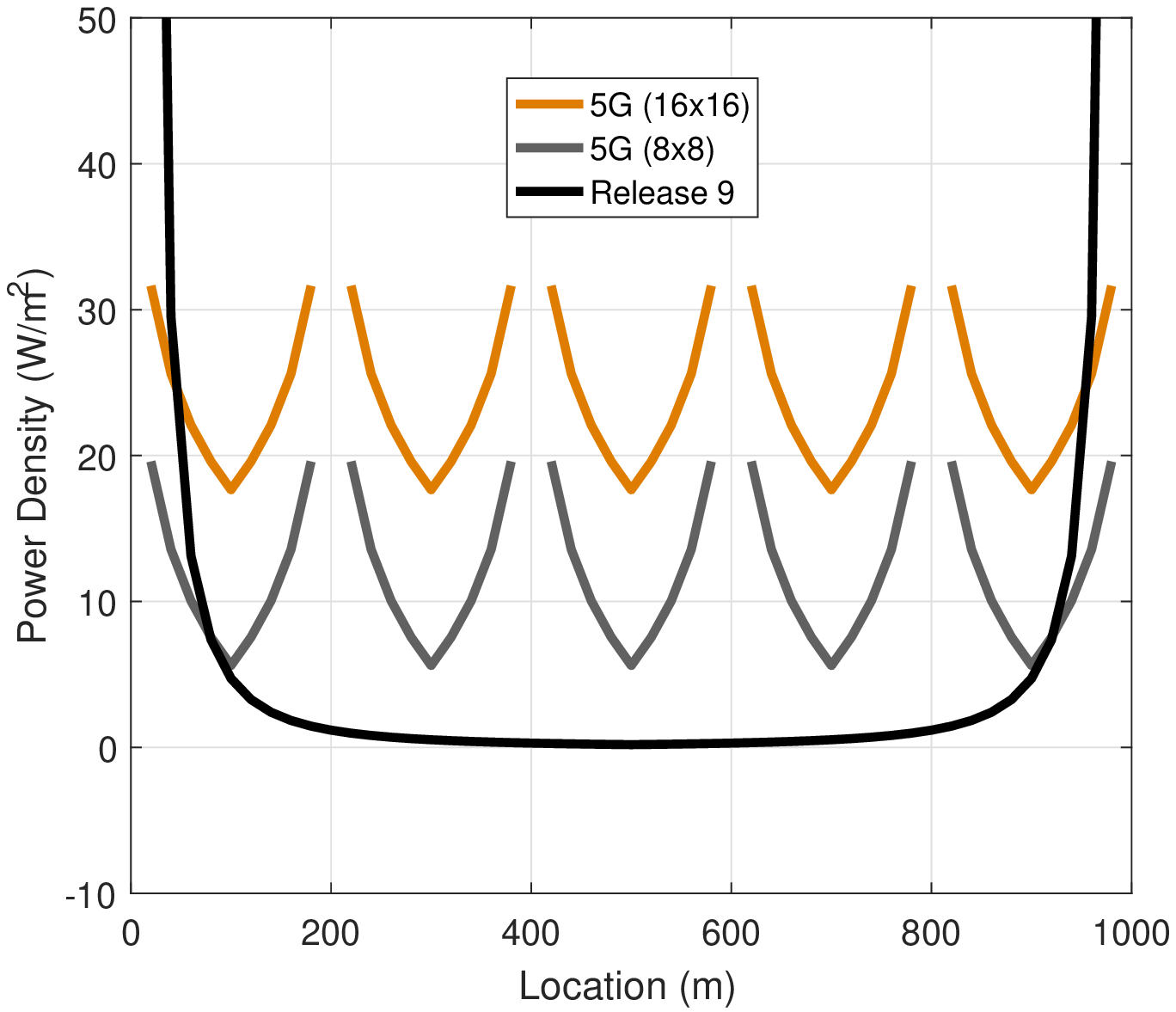}
\caption{Power density (\ref{eq_pd}) versus UE location in a 5G and Release 9 system}
\label{fig_pd}
\endminipage\hfill
\minipage{0.45\textwidth}
\centering
\includegraphics[width = \linewidth]{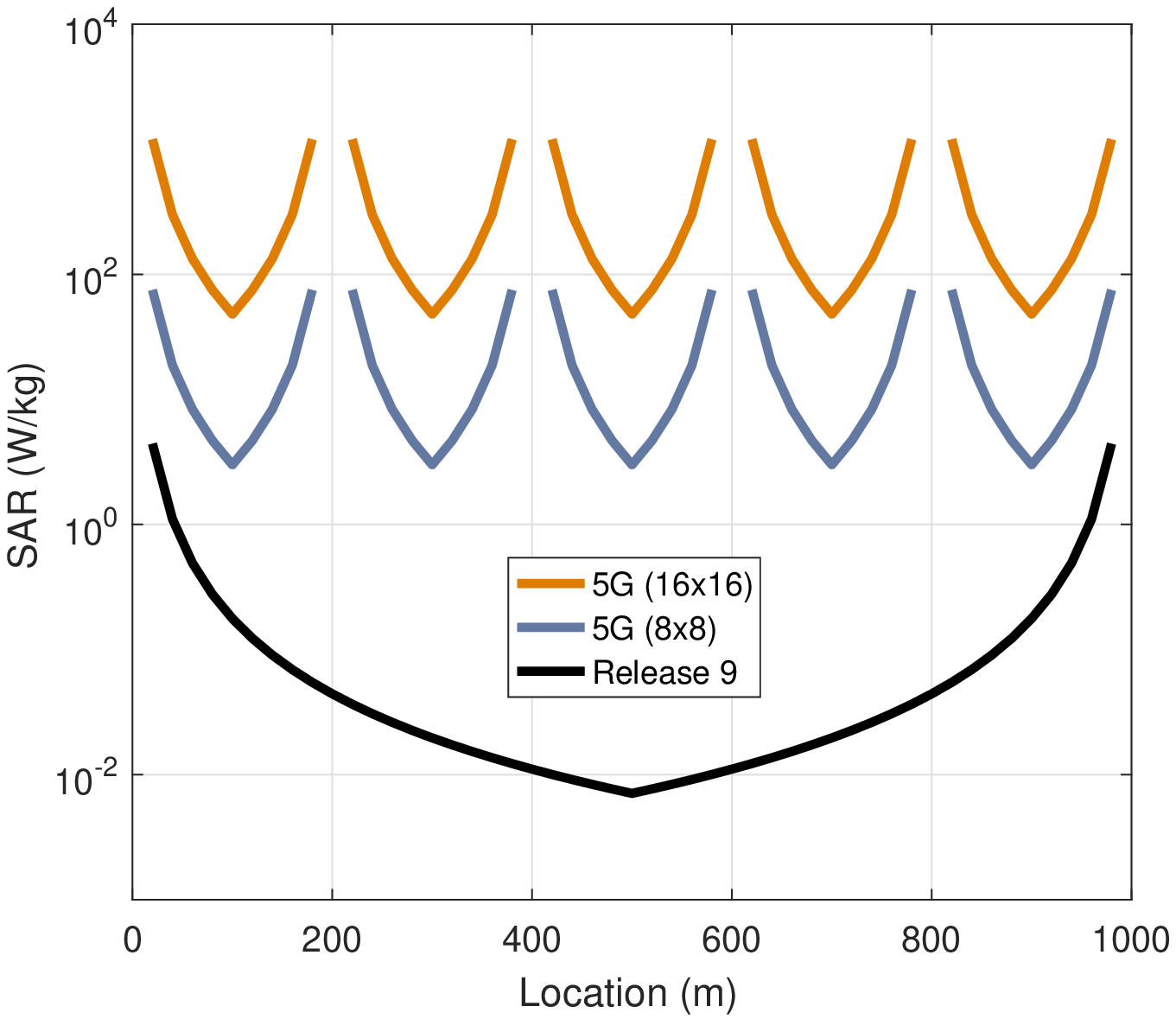}
\caption{SAR (\ref{eq_sar}) versus UE location in a 5G and Release 9 system}
\label{fig_sar}
\endminipage
\end{figure*}

Now, we can rewrite an SAR given in (\ref{eq_sar}) in terms of $d$ for calculation in a cellular communications system, which is also a function of $\phi$ \cite{gandhi86}\cite{chahat12}, as
\begin{align}\label{eq_sar_phi}
\text{SAR}\left(d\right) = \text{SAR}\left(\phi\right) = \frac{2S_i\left(\phi\right) T\left(\phi\right) m\left(\phi\right)}{\delta \rho}
\end{align}
where $T$ is the power transmission coefficient \cite{love16} and $\delta$ is the skin penetration depth (m) at 28 GHz \cite{rappaport15}. The function $m\left(\phi\right)$ \cite{love16} is dependent on the tissue properties of dielectric constant ($\epsilon^{*}$).

In order to accurately study a mmW signal propagation and absorption in a human body, investigation on the parameters related to dielectric measurements on human skin are necessary. Specifically the values of the parameters, $\rho$, $\epsilon^{*}$, $\delta$, $T$, and $m(\phi)$ are obtained from prior related work \cite{em05}\cite{rappaport15}\cite{tr38900}\cite{database14}.

\section{Evaluation of Human RF Exposure}\label{sec_results}
In this section, we analyze the results for the performance of 5G technology and make a comprehensive comparison of the model with present Release 9. First we show the performance for 5G in terms of service quality and then make a deeper interest in the health impacts due to exposure to EM emissions at mmW radiation.

\subsection{Data Rate}
We consider two antenna array sizes: 8 $\times$ 8 and 16 $\times$ 16 for 5G analysis. As we consider 3 sectors under each AP, it is adequate for each antenna to have the coverage of 120$^{\circ}$ capability to cover an entire 360$^{\circ}$ range of the cell.

Figs. \ref{fig_pr_5g} and \ref{fig_pr_release9} show the signal power received at a UE, $P_{R,ue}\left(\mathtt{x}_{ue}\right)$, at different locations in 5G and Release 9 scenarios, respectively. The most significant factor that determines a received signal power is path loss that is in turn dominated by the LoS probability provided differently in each path loss model \cite{tr38900}. The received power decreases sharply with increasing distance in both systems, but as the APs are located at much closer positions for 5G, the received power bounces back to increase again while it keeps on decreasing with increasing distance in a Release 9 system. Also, it can be seen from Figs. \ref{fig_pr_5g} and \ref{fig_pr_release9} that even at the cell edges (at 100, 300, 500, 700, and 900 m), the received power is still remarkably higher for all 5G scenarios than the respective scenarios of the Release 9. One key rationale behind this outperformance can obviously be found as the higher antenna gain that an AP can form by adopting the larger phased arrays.

Figs. \ref{fig_rate_5g} and \ref{fig_rate_release9} show data rates that can be achieved in a 5G and a Release 9 system, respectively, to represent the downlink performances. One can obviously find that a higher received power directly leads to a higher data rate (as observed from comparison to Figs. \ref{fig_pr_5g} and \ref{fig_pr_release9}), considering the data rate that is calculated from (\ref{eq_r}). Fig. \ref{fig_rate_5g} illustrates a comparison of data rates achieved in a 5G downlink system between different AP's phased array size--16 $\times$ 16 and 8 $\times$ 8. It can be seen that a UE in all 5G scenarios yields a downlink data rate above 13 Gbps even at a cell edge. Fig. \ref{fig_rate_release9} presents downlink data rates in a Release 9 system.

It should be emphasized from Figs. \ref{fig_rate_5g} and \ref{fig_rate_release9} that in spite of the disadvantage in the propagation due to the higher carrier frequency, a 5G system presents approximately 20-times higher downlink rates compared to a Release 9 system regardless of (i) the path loss model and (ii) an AP's phased array size. The main rationale behind such a significant outperformance is the smaller ISD in a 5G system. It is thus evident that the 5G mmW technology provides significantly better performance to the consumer as it provides better signal strength with higher data transmission capabilities at the user end.

\subsection{Human RF Exposure}
Now we show that even considering such shallow penetration depth due to high frequencies, a downlink RF emission causes significantly higher SAR in mmW. In this section, the PD and SAR are compared between a 5G and a Release 9 system. It still remains not concluded in the literature which of PD and SAR is more appropriate to represent the human RF exposure level in far-field RF propagations. We claim that \textit{SAR should not be excluded} in measurement of human RF exposure in mmW downlinks. The rationale is that in spite of shallower penetration into a human body compared to lower frequencies, a mmW RF field causes a higher SAR due to (i) smaller cell radius and (ii) higher concentration of RF energy per beam via adoption of larger phased array.

Fig. \ref{fig_pd} compares the PD between the downlinks of 5G and Release 9. One can find far higher PDs in 5G downlinks compared to those of a Release 9 system. The same rationale yields this higher PD in 5G downlinks: the PD in a 5G system bounces back up at a shorter distance compared to a Release 9 system due to the smaller ISD. In other words, the denser deployment of cell sites in 5G keeps PDs higher in more areas in a network than in a Release 9 network. At a distance about 50 m from the nearest AP for 5G, the user is exposed to a significant PD value when a 16 $\times$ 16 array is used. Thus, when a larger phased antenna is used or when a user moves closer to the AP, the PD value becomes a major health concern which inevitably requires more research about health effects of 5G before it is deployed successfully by strictly following the RF emission standards.

We show the comparison of SAR also between 5G and present existing scenario in Fig. \ref{fig_sar} for far-field to have a better understanding about the health impacts of RF emissions into human body. The SAR requirements for near-field is stated in \cite{wu15}, but to the best of our knowledge, there is no standard provided for SAR in far-field scenario so far as it is expected that SAR does not have a significant effect on human body in far-field. Our result in Fig. \ref{fig_sar} presents that a 5G downlink does not allow a sufficient far-field propagation due to the small-cell topology. This yields a much higher SAR level than Release 9 that adopts a larger ISD that consequently yields a longer propagation that is sufficient fall down to a low enough SAR. This is resulted from the mmW radiations, antenna beam steering effects and smart antenna characteristics of 5G architecture.

The result provided in Fig. \ref{fig_sar} has a significant implication. According to the ICNIRP guidelines \cite{icnirp98}, the maximum allowable SAR level for head and trunk is 2 W/kg and for limbs it is 4 W/kg for 10 g tissue over 6 minutes of exposure for frequencies up to 10 GHz for general public (ICNIRP and FCC \cite{fcc01} do not have SAR guidelines for mmW like 28 GHz far-field scenario yet, as it is expected to be less dangerous). But our result presented in Fig. \ref{fig_sar} shows a significant increase in SAR in 5G downlinks compared to the Release 9, even in such far-field propagations. Considering the significance of a regulatory guideline in the societal endeavor to prevent injuries from over-exposure, this paper hereby strongly urges that it is not safe enough with the PD solely being considered as a basic restriction in human RF exposure in mmW operations. Our result suggests that the SAR should also be considered as a measuring parameter even for far-field, particularly in mmW communications due to its received signal strength remaining strong at an end user.

\section{Conclusions}
This paper has highlighted the significance of human RF exposure issue in downlink of a cellular communications system. This paper measured the exposure level in terms of PD and SAR, and compared them to those calculated in the Release 9 as a representative of the current mobile communications technology. Distinguished from the prior art that studied uplinks only, this paper has found that the downlinks of a 5G also yield significantly higher levels of PD and SAR compared to a Release 9. Our results emphasized that the increase stems from two technical changes that will likely occur in 5G: (i) more APs due to deployment of smaller cells and (ii) more highly concentrated RF energy per downlink RF beam due to use of larger phased arrays.

As such, unlike the prior work, this paper claims that RF fields generated in downlinks of 5G can also be dangerous in spite of far-field propagations. Therefore, we here urge design of cellular communications and networking schemes that force an AP to avoid generation of RF fields if pointed at a human user with an angle yielding a dangerous level of PD and SAR. To this end, this paper identifies as the future work proposition of techniques that reduces human exposure to RF fields in 5G downlinks.

\end{document}